\documentclass[twocolumn]{aastex63}
\usepackage{amsmath}
\usepackage{graphicx}
\usepackage{booktabs}

\newcommand{\Phipar}{\Phi_{\rm par}}
\newcommand{\Phiobs}{\Phi_{\rm obs}}
\newcommand{\ledd}{\lambda_{\rm Edd}}
\newcommand{\mbh}{M_{\rm BH}}
\newcommand{\mgal}{M_{\rm gal}}

\newcommand{\mh}{M_{h}}
\newcommand{\msun}{M_{\odot}}
\usepackage{xcolor}

\accepted{March 10, 2021}
\shorttitle{PDE of QLF at $2\lesssim z \lesssim 6$}
\shortauthors{Kim \& Im}
\begin{document}

\title{Pure Density Evolution of the Ultraviolet Quasar Luminosity Function at $2\lesssim z \lesssim 6$}
\email{yongjungkim@pku.edu.cn, mim@astro.snu.ac.kr}
%\email{mim@astro.snu.ac.kr}

\author[0000-0003-1647-3286]{Yongjung Kim}
\altaffiliation{KIAA Fellow}
%\altaffiliation{yongjungkim@pku.edu.cn}
\affiliation{Kavli Institute for Astronomy and Astrophysics, Peking University, Beijing 100871, P. R. China}
\affiliation{SNU Astronomy Research Center, Seoul National University, 1 Gwanak-ro, Gwanak-gu, Seoul 08826, Republic of Korea}

\author[0000-0002-8537-6714]{Myungshin Im}
\affiliation{SNU Astronomy Research Center, Seoul National University, 1 Gwanak-ro, Gwanak-gu, Seoul 08826, Republic of Korea}
\affiliation{Astronomy Program, Department of Physics \& Astronomy, Seoul National University, 1 Gwanak-ro, Gwanak-gu, Seoul 08826, Republic of Korea}

%% Note that the \and command from previous versions of AASTeX is now
%% depreciated in this version as it is no longer necessary. AASTeX 
%% automatically takes care of all commas and "and"s between authors names.

%% AASTeX 6.2 has the new \collaboration and \nocollaboration commands to
%% provide the collaboration status of a group of authors. These commands 
%% can be used either before or after the list of corresponding authors. The
%% argument for \collaboration is the collaboration identifier. Authors are
%% encouraged to surround collaboration identifiers with ()s. The 
%% \nocollaboration command takes no argument and exists to indicate that
%% the nearby authors are not part of surrounding collaborations.

%% Mark off the abstract in the ``abstract'' environment. 

\begin{abstract}
Quasar luminosity function (QLF) shows the active galactic nucleus (AGN) demography as a result of the combination of the growth and the evolution of black holes, galaxies, and dark matter halos along the cosmic time.
The recent wide and deep surveys have improved the census of high-redshift quasars, making it possible to construct reliable ultraviolet (UV) QLFs at $2\lesssim z\lesssim6$ down to $M_{1450}=-23$ mag.
By parameterizing these up-to-date observed UV QLFs that are the most extensive in both luminosity and survey area coverage at a given redshift, we show that the UV QLF has a universal shape, and their evolution can be approximated by a pure density evolution (PDE).
In order to explain the observed QLF, we construct a model QLF employing the halo mass function, a number of empirical scaling relations, and the Eddington ratio distribution.
We also include the outshining of AGN over its host galaxy, which made it possible to reproduce a moderately flat shape of the faint end of the observed QLF (slope of $\sim-1.1$).
This model successfully explains the observed PDE behavior of UV QLF at $z>2$, meaning that the QLF evolution at high redshift can be understood under the framework of halo mass function evolution.
The importance of the outshining effect in our model also implies that there could be a hidden population of faint AGNs ($M_{1450}\gtrsim-24$ mag), which are buried under their host galaxy light.
\end{abstract}

%% Keywords should appear after the \end{abstract} command. 
%% See the online documentation for the full list of available subject
%% keywords and the rules for their use.
%\keywords{cosmology: observations --- galaxies: active --- galaxies: high-redshift --- 
%quasars: supermassive black holes --- surveys}

%% From the front matter, we move on to the body of the paper.
%% Sections are demarcated by \section and \subsection, respectively.
%% Observe the use of the LaTeX \label
%% command after the \subsection to give a symbolic KEY to the
%% subsection for cross-referencing in a \ref command.
%% You can use LaTeX's \ref and \label commands to keep track of
%% cross-references to sections, equations, tables, and figures.
%% That way, if you change the order of any elements, LaTeX will
%% automatically renumber them.
%%
%% We recommend that authors also use the natbib \citep
%% and \citet commands to identify citations.  The citations are
%% tied to the reference list via symbolic KEYs. The KEY corresponds
%% to the KEY in the \bibitem in the reference list below. 

\section{INTRODUCTION} \label{sec:introduction}

As an observable black hole (BH) over a wide redshift range, quasars, the most powerful active galactic nuclei (AGNs), have been playing a pivotal role in understanding the formation and the growth of BH along the cosmic time. 
The quasar demography at a given redshift, represented by quasar luminosity function (QLF), is the result of intertwining evolution of several physical properties of BHs and their host galaxies: the gas fueling mechanism, the obscuration of quasars by dust, the growth of quasar host galaxies/halos, to name a few.
Therefore, by studying the cosmic evolution of QLF, one can comprehend a general picture of how halos, galaxies, and BHs evolved together.

\begin{figure*}
\centering
\epsscale{1.2}
\plotone{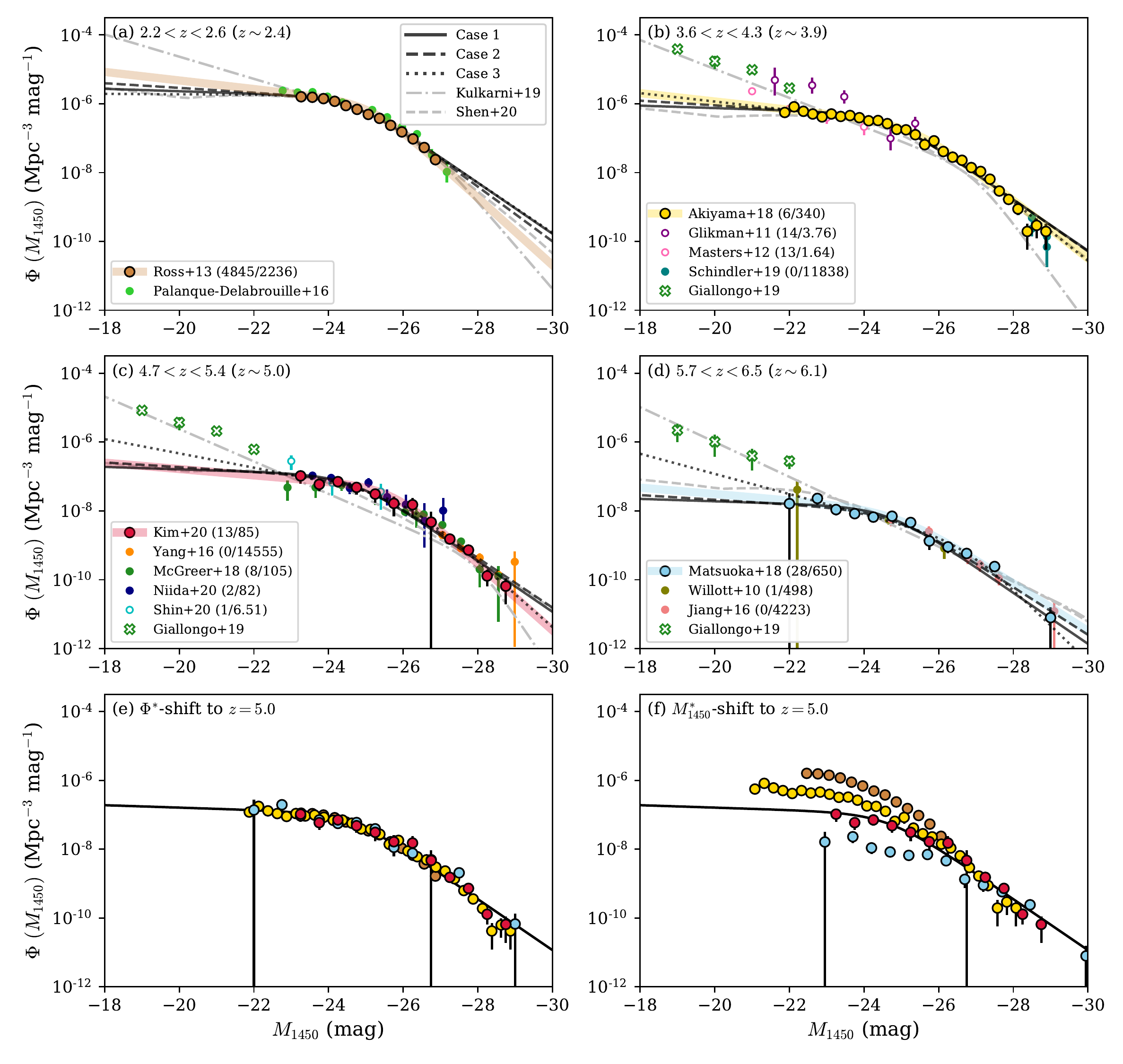}
\caption{
(a)-(d): Observed QLFs at $2\lesssim z\lesssim6$. The redshift range and the central redshift are marked in each panel.
The filled (open) circles denote the QLFs from the large (small) area surveys.
The numbers in the parentheses in the legend represent the number of spectroscopically identified quasars at $M_{1450}>-24$ mag ($N_{\rm spec,-24}$) and the survey area in deg$^{2}$ of each study, i.e., ($N_{\rm spec,-24}$/Area).
The QLFs selected for this study are highlighted by the black circles \citep{Ross13,Akiyama18,Matsuoka18,Kim20} with their parametric QLFs (thick translucent line).
The green open crosses are from the X-ray-selected quasar sample \citep{Giallongo19}.
The black solid, dashed, and dotted lines denote the best-fit QLF in cases 1, 2, and 3, respectively.
The empirical models of \cite{Kulkarni19} and \cite{Shen20} are shown as the gray dot-dashed and dashed lines, respectively.
(e)-(f): Shifted QLFs by scaling $\Phi^{*}$ or $M_{1450}^{*}$ to fit to the faint-end and bright-end of the $z\sim5$ QLF, respectively. The solid line represents the best-fit $z\sim5$ result in case 1.
\label{fig:qlf}}
\end{figure*}

The observed QLFs have been examined in various ways.
\cite{Hopkins07} and \cite{Shen20} assembled a large number of observed QLFs and determined the bolometric QLF as a function of redshift.
Similarly, \cite{Kulkarni19} also parameterized the ultraviolet (UV) QLF, using available data at that time.
Other studies focused on the determination of QLF from the empirically/observationally constrained relations among quasars, galaxies, and dark matter halos, by using conditional luminosity function \citep{Conroy13,Ren20} or continuity equation \citep{Tucci17}.
\cite{Veale14} also presented simple models with the growth-based evolution of BHs and galaxies. 
These studies show that the QLF evolution at $z\lesssim3$ is complicated, favoring an interwoven evolution of the number density, the luminosity, and the bright/faint-end slopes.

At $z>3$, however, the QLF studies had been fundamentally hampered by the lack of faint quasars that can define the QLF faint-end slope.
For example, the Sloan Digital Sky Survey (SDSS) discovered quasars up to $z\sim6$, but the SDSS high-redshift quasar sample is limited to the brightest ones with the absolute magnitude at $1450$ \AA~of $M_{1450}<-25$ mag \citep{Jiang16}.
With bright quasars alone, the previous studies were limited to constraining the bright-end slope of QLFs \citep{Hopkins07,Conroy13,Veale14}.
Very recently, subsequent large-area imaging surveys expanded the luminosity range of $z\gtrsim4$ quasar sample down to $M_{1450}=-23$ mag or fainter \citep{Akiyama18,Matsuoka18,McGreer18,Kim19,Kim20,Niida20}.
Using a sample of dozens or more of faint quasars, one can now obtain a meaningful constraint on the faint-end of UV QLF.

In this letter, we investigate the evolution of QLF at $2\lesssim z\lesssim6$ using the most up-to-date observed QLFs.
We show that the UV QLF evolution is dominated by a pure density evolution (PDE) and provide possible interpretations for this rather unexpected result.
The cosmological parameters we adopted are $\Omega_{m}=0.3$, $\Omega_{\Lambda}=0.7$, and $H_{0}=70$ km s$^{-1}$ Mpc$^{-1}$.

\section{Observed UV QLF\label{sec:obsqlf}}

The step towards figuring out the evolutionary trend of quasar demography is to select the QLFs that are least biased at their faint ends. 
Among dozens of UV QLFs at $2\lesssim z\lesssim6$ in the literature, we first collect the QLFs based on the quasars selected by their unique rest-UV colors, rather than by X-ray detection (e.g., \citealt{Giallongo19}) or photometric variability (e.g., \citealt{Palanque16}).
Then, we excluded the QLFs derived from the quasar sample from a small survey area ($<10$ deg$^{2}$) or consisting of only bright ones ($M_{1450}<-24$ mag).
If there are several results at a given redshift, we chose the one that used the largest number of spectroscopically identified quasars at $M_{1450}>-24$ mag ($N_{\rm spec,-24}$).
For example, at $z\sim5$, there are three comparable studies: \citeauthor{McGreer18} (\citeyear{McGreer18}; 105 deg$^{2}$), \citeauthor{Kim20} (\citeyear{Kim20}; 85 deg$^{2}$), and \citeauthor{Niida20} (\citeyear{Niida20}; 82 deg$^{2}$).
But their $N_{\rm spec,-24}$ are 8, 13, and 2, respectively, so we took the result of \cite{Kim20}.

Figure \ref{fig:qlf} shows the selected QLFs (marked with black circles) at four different redshift bins \citep{Ross13,Akiyama18,Matsuoka18,Kim20} of which central redshifts are $z\sim2.4$, 3.9, 5.0, and 6.1.
These QLFs are shifted in absolute magnitudes and number densities to our chosen cosmological parameters.
For the \cite{Ross13} QLF at $z\sim2.4$, their $M_i(z=2)$ magnitudes are converted to $M_{1450}$ following the prescription in Appendix B of their work.
\cite{Akiyama18} used the photometric redshift sample with only $N_{\rm spec,-24}=6$.
However, their photometric redshift accuracy is small enough ($\Delta z/(1+z)\sim 0.03$) to accurately trace the LF shape.
We note that the bright ends ($M_{1450}\lesssim-27$) of the three selected QLFs \citep{Akiyama18,Matsuoka18,Kim20} are determined by the bright quasar sample from other studies (e.g., SDSS).
As a result, the selected QLFs cover wide ranges in luminosity ($-30<M_{1450}<-23$) and survey area ($> 80$ deg$^{2}$), and can be considered the best-determined QLF to date.

In Figure \ref{fig:qlf}, we compare QLFs from different literature. After homogenizing the cosmological parameters to our chosen values, QLFs are shifted in number density to the central redshift of the selected QLFs within each redshift panel, with a number density scaling factor we present in this work (case 1 in Section \ref{sec:shape}).
The selected QLFs are in good agreement with the other QLFs based on large-area survey data (filled circles; \citealt{Willott10,Jiang16,Palanque16,Yang16,McGreer18,Schindler19,Niida20}), although some QLFs from small area surveys ($<10$ deg$^{2}$) tend to deviate from the selected ones (open circles and crosses; \citealt{Glikman11,Giallongo19}).

The thick translucent lines in Figure \ref{fig:qlf} denote the parametric QLF ($\Phipar$), canonically described by a double power-law (DPL) function:

\begin{equation}
\begin{aligned}
&\Phipar(M_{1450},z)=\\
&\frac{\Phi^{*}}{10^{0.4(\alpha+1)(M_{1450}-M_{1450}^{*})}+10^{0.4(\beta+1)(M_{1450}-M_{1450}^{*})}},\label{eq:Phipar}
\end{aligned}
\end{equation} 

\noindent where $\Phi^{*}$ is the normalization factor, $M_{1450}^{*}$ is the break magnitude, and $\alpha$ and $\beta$ are the faint- and bright-end slopes, respectively.
The best-fit parameters are taken from the corresponding references \citep{Ross13,Akiyama18,Matsuoka18,Kim20}.

The striking feature in Figure \ref{fig:qlf} is the similarity of the shape of QLFs at different redshifts.
In the panels (e) and (f), we show QLFs shifted only in density and only in luminosity, respectively.
We note that the density-shift alone makes the QLFs overlap almost perfectly with each other.

\section{Redshift Evolution of QLF\label{sec:shape}}

To describe the redshift evolution, we assume polynomial functions for the four parameters of $\Phipar$:

\begin{equation}
X(z) = \sum_{i=0}^{n_{X}} C_{X,i}~(z-z_p)^{i}\label{equ:poly},
\end{equation}

\noindent where $X\in\{\log_{10}\Phi^{*},~M_{1450},~\alpha,~\beta\}$, $C_{X,i}$ is the $i$-th order coefficient for the parameter $X$, $n_X$ is the maximum order, and $z_p=2.2$ (pivot redshift).
Here we consider three cases:

\begin{itemize}
\item Case 1: A PDE model where only $\Phi^{*}$ evolves, and to the 2nd order, i.e.,  $n_X\in\{2,0,0,0\}$ for $X$.

\item Case 2: In addition to the number density evolution of Case 1, we allow the bright-end slope $\beta$ to evolve but to the 1st order (see Figure \ref{fig:qlf}), i.e., $n_X\in\{2,0,0,1\}$ for $X$.

\item Case 3: In addition to the number density evolution of Case 1, we allow all the other parameters to evolve but to the 1st order, i.e., $n_X\in\{2,1,1,1\}$ for $X$.
\end{itemize}

We fit these functions to the observed QLF data points ($\Phiobs$), with the maximum likelihood estimation.
For this, we used the \texttt{emcee} Python package\footnote{\url{https://emcee.readthedocs.io/en/stable/}} \citep{Foreman13} for the Markov chain Monte Carlo sampling of the DPL parameters.
We used a likelihood function of $\mathcal{L}=-\frac{1}{2}\sum\left[(\Phiobs-\Phipar)/\sigma_{\Phiobs}\right]^{2}$, where $\sigma_{\Phiobs}$ is the $1\sigma$ uncertainty of $\Phiobs$ from the literature.
We used uninformative priors on the parameters within the reasonable ranges: $-10<\log_{10}\Phi^{*}<0$, $-30<M_{1450}^{*}<-23$, $-5<\alpha,\beta<0$, and $-1<C_{X,i}<1$.
The best-fit results with 1$\sigma$ errors are taken to be the median values with standard deviations of their posterior distributions with 10,000 chains, listed in Table \ref{tbl:params}. 

The resultant QLFs are shown in Figure \ref{fig:qlf} with the solid, dashed, and dotted lines.
They are consistent with each other, supported by the fact that the residuals ($\Delta\Phi=\log_{10}\Phiobs-\log_{10}\Phipar$) for the selected QLFs have the normal-like distributions with a standard deviation of only 0.15 to 0.12 dex from cases 1 to 3.
We note that the reduced chi-square ($\chi^{2}_{\nu}$) value between the best-fit result and observation naturally decreases as the model becomes complicated with the increasing number of free parameters, but only mildly; $\chi^{2}_{\nu}=3.14$, 2.93, 2.22 from cases 1 to 3.

\begin{figure*}
\centering
\epsscale{1.2}
\plotone{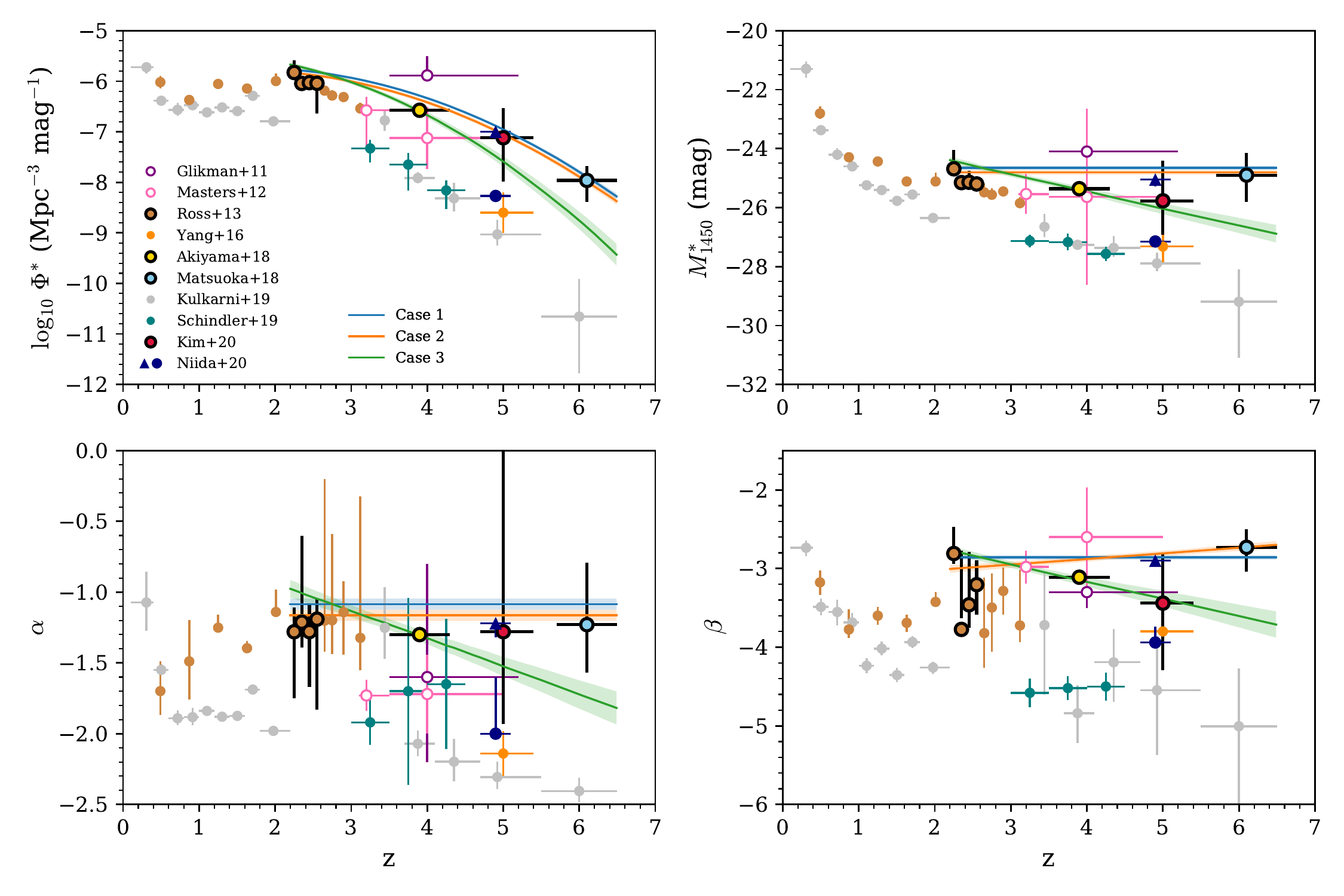}
\caption{
QLF parameters along the redshift.
The blue, orange and green lines represent the best-fit results of cases 1, 2, and 3, respectively, with their 1$\sigma$ uncertainties (shaded regions).
The best-fit parameters from the literature, which are determined without any fixed parameters, are shown as the same symbols in Figure \ref{fig:qlf}, while the data points from the QLFs selected for our model fitting are highlighted by black circles.
For \cite{Niida20}, we plot both their results with fixed $\beta=-2.9$ (navy triangles) and without any fixed parameters (navy circles).
\label{fig:params}}
\end{figure*}

In Figure \ref{fig:params}, we show the changes in parameters of our models along the redshift with the best-fit parameters of $\Phipar$ in literature.
Note that we only plot the results determined from the maximum-likelihood method to individual quasars (not to the binned QLFs).
Our results are broadly consistent with the parameters of the observed QLFs selected for the study.
Unlike cases 1 and 2, there are large discrepancies between the case 3 fit and the local best-fit values at $z\sim6$, although the largest number of free parameters were used in case 3.
This is due to the high dependence of our MCMC run on the lower-redshift QLFs \citep{Ross13,Akiyama18} that have a larger number of data points with smaller uncertainties than the high-redshift QLFs.
The case 3 result is the best mathematically, but we point out that it is only slightly more accurate than others in terms of $\Delta \Phi$ and $\chi^{2}_{\nu}$.

\begin{deluxetable}{lccc}
\tabletypesize{\scriptsize}
\tablecaption{Best-fit DPL Parameters\label{tbl:params}}
\tablewidth{0pt}
\tablehead{
\colhead{$X$} & \colhead{$C_{X,0}$} & \colhead{$C_{X,1}$} & \colhead{$C_{X,2}$}
}
\startdata
\multicolumn{4}{c}{Case 1}\\
$\log_{10}\Phi^{*}$ 	& $-5.77\pm0.03$ 	& $-0.12\pm0.02$ & $-0.11\pm0.01$ \\
$M^{*}_{1450}$ 	& $-24.64\pm0.07$ 	& ... & ... \\
$\alpha$ 			& $-1.09\pm0.04$ 	& ... & ... \\
$\beta$ 			& $-2.86\pm0.03$ 	& ... & ... \\
\hline
\multicolumn{4}{c}{Case 2}\\
$\log_{10}\Phi^{*}$ 	& $-5.83\pm0.03$ 	& $-0.13\pm0.02$ & $-0.11\pm0.01$ \\
$M^{*}_{1450}$ 	& $-24.82\pm0.08$ 	& ... & ... \\
$\alpha$ 			& $-1.17\pm0.04$ 	& ... & ... \\
$\beta$ 			& $-3.01\pm0.05$ 	& $0.07\pm0.02$ & ... \\
\hline
\multicolumn{4}{c}{Case 3}\\
$\log_{10}\Phi^{*}$ 	& $-5.68\pm0.04$ 	& $-0.32\pm0.03$ & $-0.13\pm0.01$ \\
$M^{*}_{1450}$ 	& $-24.40\pm0.11$ 	& $-0.59\pm0.09$ & ... \\
$\alpha$ 			& $-0.98\pm0.07$ 	& $-0.19\pm0.04$ & ... \\
$\beta$ 			& $-2.78\pm0.06$ 	& $-0.22\pm0.05$ & ... \\
\enddata
\tablecomments{$\Phi^{*}$ is in units of Mpc$^{-3}$ mag$^{-1}$. }
\end{deluxetable}

There are several QLF parameters in disagreement \citep{Yang16,Kulkarni19,Schindler19,Niida20}.
In the case of \cite{Niida20}, their main result with a fixed slope of $\beta=-2.9$ (navy triangles) always shows higher values in all the parameters than those without any fixed parameters (navy circles).
This implies that the correlation between QLF parameters may show up as mathematical difference (e.g., see \citealt{Matsuoka18,Kim20}).
Also, the results biased toward lower values than our models \citep{Yang16,Kulkarni19,Schindler19} can be attributed to the use of QLFs that are not sufficiently constrained due to small-number statistics and/or faint-end incompleteness.

Overall, we conclude that, under the current number of datasets, the QLF at $z>2$ can be well described with a PDE model (case 1) and the addition of the evolution in the other QLF parameters does not improve the fitting result significantly.

The QLF evolution has been studied previously, and several authors deduced the evolution models that are more complicated than the PDE scenario as presented here \citep{Hopkins07,Kulkarni19}.

But the high-redshift QLFs in \cite{Hopkins07} do not extend deep enough to reliably constrain the faint-end QLF shape.
Meanwhile, \cite{Kulkarni19} use QLFs down to $M_{1450}\sim-22$ mag at $z\gtrsim3$, but their analysis included datasets that we reject in this work due to small survey area \citep{Glikman11} or a small number of faint quasars \citep{Willott10,Kashikawa15}.
We suggest that these are the reason for the discrepancy between the previous results and our result for the QLF evolution.
In fact, the most recent work by \cite{Shen20}, including quasars over a wide magnitude range, shows a result in line with our simple PDE models at various redshifts (the gray dashed line in Figure \ref{fig:qlf}).
\cite{Niida20} also suggest little evolution in $\alpha$ and $M_{1450}$ at $4\lesssim z\lesssim6$.

\section{SIMPLE MODEL FOR QLF\label{sec:empqlf}}

The PDE trend of the QLF at $z\gtrsim2$ is intriguing since, previously, QLFs have been depicted to evolve in a much more complicated way.
To explain the universal QLF shape and the PDE behavior, we constructed a theoretically motivated and empirically calibrated QLF model.

Recent observations for individual quasars suggest that there are only small or negligible changes in their intrinsic properties at $2<z<6$:  the Eddington ratio ($\ledd$) distribution \citep{Mazzucchelli17,Kim19,Onoue19,Shen19}, the obscured fractions \citep{Vito18}, the BH-to-galaxy mass ratios \citep{Izumi19}, and the metal enrichments \citep{Shin19,Schindler20}.
This implies that the QLF is not determined by the difference in the characteristics of quasars at $z>2$, but rather by the characteristics of galaxies/halos in which quasars are embedded.
Here, we construct a simple model that is built on a halo mass ($\mh$) function and several scaling relations to see how such a model can reproduce the observed QLF shape and the PDE behavior.

Starting with the $\mh$ function of \cite{Jenkins01}, we converted this to the stellar mass ($\mgal$) function, using the $\mh$-$\mgal$ relation of \cite{Behroozi19} with a scatter of $0.3-0.025\times\log(\mh/10^{10}~\msun)$ dex, inferred from their Figure 12.
We used the $z=2$ relation of all galaxies as a reference, considering the broadly constant shape of their $\mh$-$\mgal$ relation at a redshift range of $2\leq z\leq6$.

Second, the $\mgal$ function was converted to the BH mass ($\mbh$) function, following the \cite{Kormendy13} relation with a scatter of 0.4 dex.
Since their relation is given for bulge mass, we used the $\mgal$-dependent bulge-to-total mass ratio, $B/T(\mgal)=\min[1,~10^{-3.35}\mgal^{0.29}]$, derived from a sample of \cite{Mendel14}, giving $\mbh\propto\mgal^{1.5}$.

Third, the $M_{\rm BH}$ function was converted to the bolometric QLF by convolving it with a log-normal\footnote{We also tested the function in the form of Schechter function, but there is no significant difference in the resultant QLF shape as in \cite{Veale14}. While the physical interpretation could be different, we only considered the light-bulb scenario of quasars for simplicity.} Eddington ratio distribution,

\begin{equation}
P(\ledd)=\frac{P^*}{\sigma\sqrt{2\pi}}\exp\left(\frac{-(\log\ledd-\log\lambda_{\rm Edd}^{*})^2}{2\sigma^{2}}\right)\label{equ:ln},
\end{equation}

\noindent where $P^*$ is the normalization factor related to the \emph{observable} duty cycle related to quasar lifetime and UV obscuration.
We set $\log\ledd^{*} = -0.5$ and $\sigma=0.3$ dex \citep{Shen19}.
We converted the bolometric luminosity to $M_{1450}$ using the correction factor from \cite{Shen20}.

Lastly, we consider the outshining of AGN in quasar hosts.
Previous high-redshift quasar surveys introduced the point-source selection, so it can be assumed that a BH needs to outshine its host galaxy to be selected as a quasar \citep{Ni19}.
We classified an AGN as a quasar only when its $M_{1450}$ is twice brighter than the UV magnitude of its galaxy at $1500$ \AA~($M_{\rm1500,gal}$) inferred from its $\mgal$ at a given redshift \citep{Behroozi19}.

Our QLF model has a number of adjustable parameters, but we allowed only one parameter, $P^{*}$ in equation (\ref{equ:ln}), to change its value as a function of redshifts.
We scaled $P^*$ to maximize the likelihood function between the model QLF and the selected QLFs at each redshift, and the results are discussed in the following section.

\begin{figure}
\centering
\epsscale{1.2}
\plotone{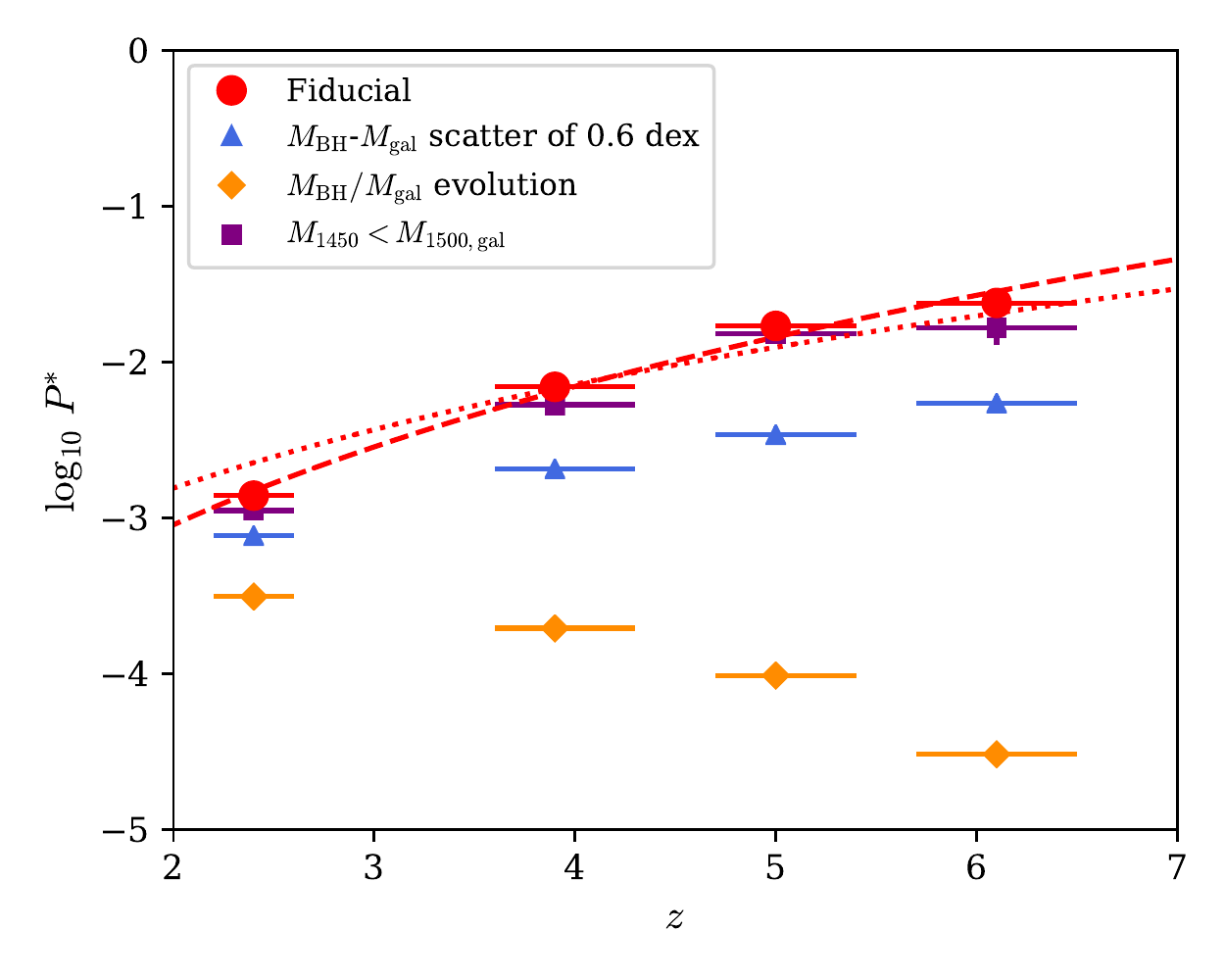}
\caption{
The duty cycle parameter $P^{*}$ along the redshift.
The red circles represent our fiducial model, with $P^{*}$ changing roughly as $(1+z)^{3}$ (dotted line) or $(1+z)^{4}$ (dashed line).
The orange diamonds, blue triangles, and purple squares denote the results when including the redshift evolution of $\mbh/\mgal$, increasing the scatter in the $\mbh$-$\mgal$ relation to 0.6 dex, or adopting a different cut for outshining effect, respectively.
\label{fig:logP}}
\end{figure}

\section{Results \& Discussion\label{sec:discussion}}

\begin{figure*}
\centering
\epsscale{1.2}
\plotone{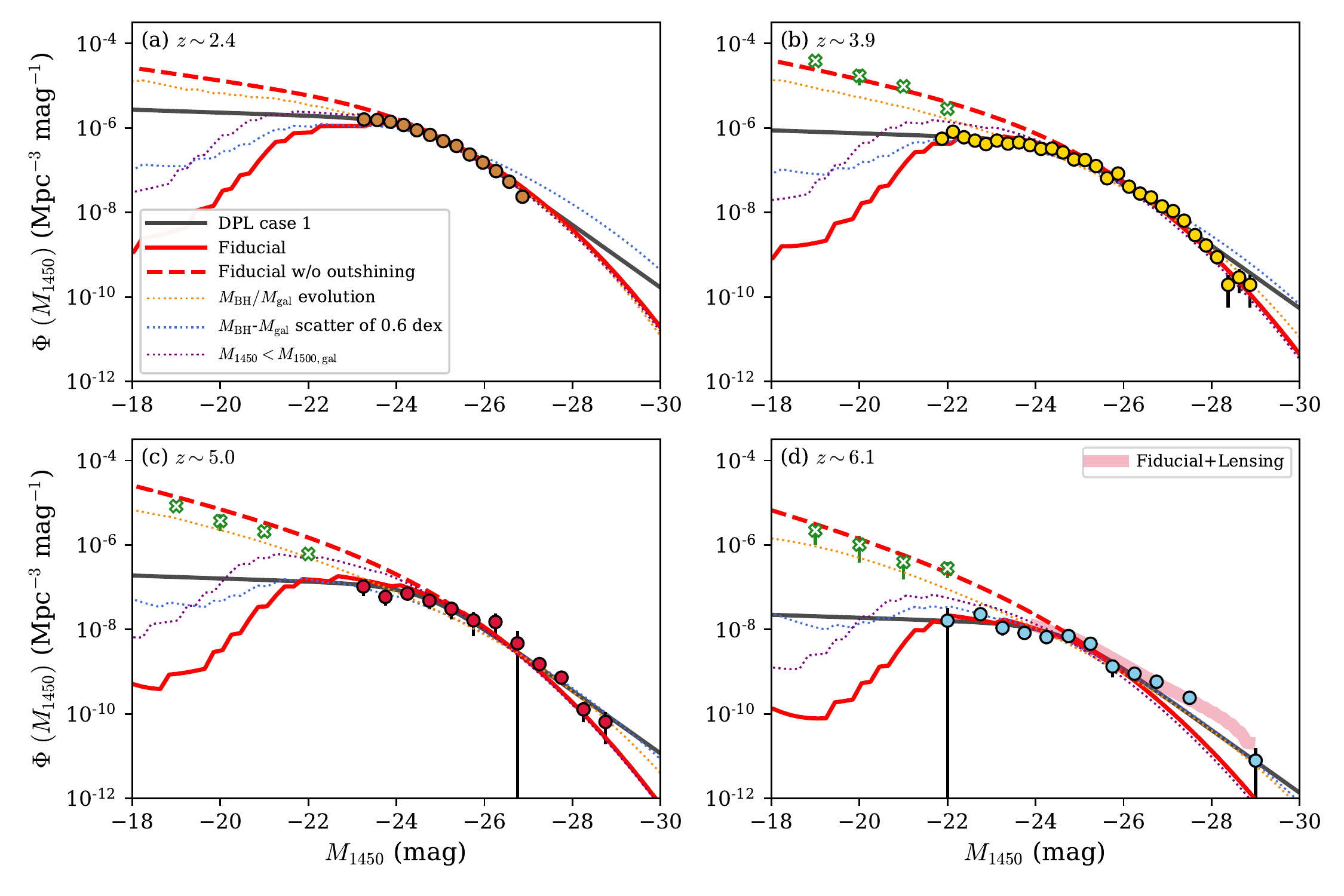}
\caption{
Comparison of the observed QLFs (points), the case 1 QLFs (black lines), and the model QLFs (red lines) at various redshifts.
The red solid (dashed) lines represent the model QLF with (without) the outshining effect.
The observed QLFs are given in the same symbol as in Figure \ref{fig:qlf}.
The orange, blue, and purple dotted lines denote the results with the changes in assumptions, similar to Figure \ref{fig:logP}.
The red translucent line in the panel (d) is the expected QLF boosted by the lensing effect assuming the intrinsic QLF slope of $\beta=-3.7$ \citep{Pacucci20}.
\label{fig:model}}
\end{figure*}

The resultant $P^*$ values are shown as the red circles in Figure \ref{fig:logP}.
Their 1$\sigma$ errors (68\%) were calculated from 100 mock QLFs generated by adding random errors to the $\Phiobs$ points.
The change in $P^{*}$ is roughly proportional to $(1+z)^{3}$ or $(1+z)^{4}$, shown as the red dotted and dashed lines, respectively.
This could be related to the cubic evolution of gas density along the redshift, but we caution that $P^{*}$ can evolve in a very different way, when some of the model assumptions are modified (see below).

Figure \ref{fig:model} shows that the model QLFs (red solid lines) agree well with the observed ones.
Like in the reproduction of galaxy luminosity function from a halo mass function, modeling a QLF from a halo mass function has a tendency of overproducing the number density at low and high luminosity ends.
In our model, the faint end of the QLF is suppressed mainly by the outshining effect.
If there is no outshining effect, as represented by the red dotted lines in Figure \ref{fig:model}, the  model would overproduce the faint QSO population.
At the bright end, we succeeded in matching the observed QLF by introducing scatters in the scaling relations and the Eddington ratio distribution.

The successful suppression of the QLF faint-end via the outshining effect suggests that there may be a large number of faint AGNs ($M_{1450}\gtrsim-24$ mag) that are not identified as quasars due simply to their faintness with respect to host galaxies (the dotted lines in Figure \ref{fig:model}).
This is also in line with the recent claim on the rapid decrease of AGN fraction among UV sources at that magnitude range \citep{Bowler21}.
Some of the QLFs based on the X-ray detection suggests a large number of faint AGNs, which has been a subject of controversy \citep{Giallongo19}.
The faint AGNs, outshone by their host galaxies in UV (the red dashed lines in Figure \ref{fig:model}), may explain such a large number of X-ray faint AGNs by some studies (the green crosses in Figure \ref{fig:model}).
We also note that the X-ray QLF, converted from our model without outshining effect using the correction factor from \cite{Shen20}, is roughly consistent with those from the recent X-ray observations at $z\sim4$ \citep{Aird15,Vito18}.
Furthermore, our outshining model suggests that the X-ray QLF does not need to follow the PDE behavior of UV QLF.
If this is true and the UV photon escape fraction of the faint AGNs is as high as luminous quasars, the faint AGN population may be responsible for a large fraction of the UV photons required for ionizing the intergalactic medium \citep{Madau15,Giallongo19}. 
For example, our model without outshining effect gives an ionizing emissivity at $912$ \AA~of $\epsilon_{912}\sim10^{24}$ erg s$^{-1}$ Hz$^{-1}$ Mpc$^{-3}$ at $z=6$, using the equations (5) and (6) of \cite{Kim20}, which is an order of magnitude higher than the value we would get from the observed QLF \citep{Matsuoka18}.

Our fiducial model does not include the evolution in the $\mbh$-$\mgal$ scaling relation, although several studies support the scaling relation evolution.
The scaling relation evolution states that $\mbh$ for a given $\mgal$ increases as a function of redshift (i.e., the BH grows first, followed by the galaxy growth).
Therefore, the quasar luminosity increases for a given $\mgal$, resulting in an overall shift of the model QLF toward higher luminosity, and this is more so at higher redshifts.
Additionally, the faint end of QLF increases since the $\mbh$-$\mgal$ scaling relation is not linear in our model.
If the scaling relation evolves as $\log(\mbh/\mgal)=0.28z-2.91$ \citep{Decarli10}, however, the mismatch in the faint end is pronounced (the orange dotted lines in Figure \ref{fig:model}).
Moreover, to match the observed QLF, we need to decrease $P^{*}$ as a function of redshift (the orange diamonds in Figure \ref{fig:logP}), which seems contradictory to the expectation that $P^{*}$ stays constant or increase with redshifts.

The model QLF  under-predicts the bright end of the $z=6$ QLF.
One way to cure this problem is to increase the $\mbh$-$\mgal$ relation scatter to 0.6 dex (the blue dotted lines in Figure \ref{fig:model}), which can happen in reality since massive halos are not mandatory for extremely bright/massive quasars (e.g., \citealt{DiMatteo17,Yoon19}) and outliers from low mass halos can contaminate the bright end easily.
However, adopting this assumption overproduces the bright end of lower redshift QLFs.
If the $\mbh$-$\mgal$ relation scatter increases with redshift, such an overproduction can be solved.
Another way to solve the bright-end problem of the $z=6$ model QLF is to introduce the gravitational lensing effect.
Our model $z=6$ QLF has an intrinsic bright-end slope of $\beta\sim-3.7$, and introducing gravitational lensing effect boosts the QLF shape at the bright end to the observed numbers (the red translucent line in Figure \ref{fig:model}, taken from \citealt{Pacucci20}).
For this to be true, a significant fraction of known $z\sim6$ bright quasars must be lensed, but such lensed bright quasars are still rare \citep{Fan19,Fujimoto20}.

We also explored how the model QLF changes if we adjust the outshining effect criterion.
If we loosen the criterion to $M_{1450}<M_{\rm 1500,gal}$, to include galaxies with a bit less luminous AGN than our base assumption, the number of quasars increases slightly and mildly at the faint end (the purple line in Figure \ref{fig:model}).

\section{Conclusion\label{sec:conclusion}}

We investigated the evolution of the UV QLFs at $2\lesssim z\lesssim6$ that are complied from recent large-area surveys.
We find that the QLF evolution can be described well with PDE.
This result is somewhat unexpected in comparison to the QLF evolution at lower redshifts for which more complicated evolutionary behaviors have been found.
Furthermore, we find that the UV QLF at $z>2$ has a universal DPL function form with a faint-end slope of $\alpha\simeq-1.1$, a break absolute magnitude of $M_{1450}^{*}\simeq-24.6$ mag, and a bright-end slope of $\beta\simeq-2.9$.

To understand the universal shape of the UV QLF and its PDE, we constructed a model QLF, starting from the $\mh$ function and applying several scaling relations that connect $\mh$ to $\mgal$ and $\mbh$, and then to quasar luminosity.
Additionally, we added the outshining effect of AGN over its host galaxy.
With these ingredients, we find that our model QLF can reproduce the observed QLFs at $z>2$.
Although there may be other ways to reproduce the observed QLF behavior, we suggest that the outshining can be an important factor in shaping the UV QLF at high redshift, especially at $z\gtrsim5$.
The importance of the outshining effect implies the existence of many faint AGNs that are buried under the galaxy light, and such faint AGNs could provide a large portion of the UV photons required for ionizing the intergalactic background.
Deep and wide NIR spectroscopic surveys with future facilities could reveal such hidden populations of faint AGN, and allow us to investigate the QLF evolution in a broader context than the simple PDE scenario presented in this work.

\acknowledgments

We thank Linhua Jiang for insightful comments, and Woncheol Jang, Byungwon Kim, and Sungkyu Jung for useful discussion about statistical tests of the QLF evolution.
We thank Fabio Pacucci for providing the expected QLF by gravitational lensing effect in \cite{Pacucci20}.
This work was supported by
the National Research Foundation of Korea (NRF) grant (2020R1A2C3011091) funded by the Korean government (MSIP).
Y. K. acknowledges the support from the China Postdoc Science General (2020M670022) and Special (2020T130018) Grants funded by the China Postdoctoral Science Foundation.

%% This command is needed to show the entire author+affilation list when
%% the collaboration and author truncation commands are used.  It has to
%% go at the end of the manuscript.
%\allauthors

%% Include this line if you are using the \added, \replaced, \deleted
%% commands to see a summary list of all changes at the end of the article.
%\listofchanges


\begin{thebibliography}{}

\bibitem[Aird et al.(2015)]{Aird15} Aird, J., Coil, A.~L., Georgakakis, A., et al.\ 2015, \mnras, 451, 1892

\bibitem[Akiyama et al.(2018)]{Akiyama18} Akiyama, M., He, W., Ikeda, H., et al.\ 2018, \pasj, 70, S34


\bibitem[Ba{\~n}ados et al.(2016)]{Banados16} Ba{\~n}ados, E., 
Venemans, B.~P., Decarli, R., et al.\ 2016, \apjs, 227, 11 

\bibitem[Behroozi et al.(2019)]{Behroozi19} Behroozi, P., Wechsler, R.~H., Hearin, A.~P., et al.\ 2019, \mnras, 488, 3143

\bibitem[Bowler et al.(2021)]{Bowler21} Bowler, R.~A.~A., Adams, N.~J., Jarvis, M.~J., et al.\ 2021, \mnras, 502, 662

\bibitem[Conroy \& White(2013)]{Conroy13} Conroy, C., \& White, M.\ 2013, \apj, 762, 70



\bibitem[Decarli et al.(2010)]{Decarli10} Decarli, R., Falomo, R., Treves, A., et al.\ 2010, \mnras, 402, 2453

\bibitem[Di Matteo et al.(2017)]{DiMatteo17} Di Matteo, T., Croft, R.~A.~C., Feng, Y., et al.\ 2017, \mnras, 467, 4243

\bibitem[Fan et al.(2019)]{Fan19} Fan, X., Wang, F., Yang, J., et al.\ 2019, \apjl, 870, L11

\bibitem[Foreman-Mackey et al.(2013)]{Foreman13} Foreman-Mackey, D., Hogg, D.~W., Lang, D., et al.\ 2013, \pasp, 125, 306

\bibitem[Fujimoto et al.(2020)]{Fujimoto20} Fujimoto, S., Oguri, M., Nagao, T., et al.\ 2020, \apj, 891, 64

\bibitem[Giallongo et al.(2019)]{Giallongo19} Giallongo, E., Grazian, A., Fiore, F., et al.\ 2019, \apj, 884, 19

\bibitem[Glikman et al.(2011)]{Glikman11} Glikman, E., Djorgovski, S.~G., Stern, D., et al.\ 2011, \apjl, 728, L26

\bibitem[Hopkins et al.(2007)]{Hopkins07} Hopkins, P.~F., Richards, G.~T., \& Hernquist, L.\ 2007, \apj, 654, 731

\bibitem[Izumi et al.(2019)]{Izumi19} Izumi, T., Onoue, M., Matsuoka, Y., et al.\ 2019, \pasj, 71, 111

\bibitem[Jenkins et al.(2001)]{Jenkins01} Jenkins, A., Frenk, C.~S., White, S.~D.~M., et al.\ 2001, \mnras, 321, 372

\bibitem[Jiang et al.(2016)]{Jiang16} Jiang, L., 
McGreer, I.~D., Fan, X., et al.\ 2016, \apj, 833, 222 

\bibitem[Kashikawa et al.(2015)]{Kashikawa15} Kashikawa, N., 
Ishizaki, Y., Willott, C.~J., et al.\ 2015, \apj, 798, 28 

\bibitem[Kim et al.(2018)]{Kim18} Kim, Y., Im, M.,
Jeon, Y., et al.\ 2018, \apj, 855, 138

\bibitem[Kim et al.(2019)]{Kim19} Kim, Y., Im, M., Jeon, Y., et al.\ 2019, \apj, 870, 86

\bibitem[Kim et al.(2020)]{Kim20} Kim, Y., Im, M., Jeon, Y., et al.\ 2020, \apj, 904, 111

\bibitem[Kormendy \& Ho(2013)]{Kormendy13} Kormendy, J. \& Ho, L.~C.\ 2013, \araa, 51, 511

\bibitem[Kulkarni et al.(2019)]{Kulkarni19} Kulkarni, G., Worseck, G., \& Hennawi, J.~F.\ 2019, \mnras, 488, 1035

\bibitem[Madau \& Haardt(2015)]{Madau15} Madau, P., \& Haardt, F.\ 2015, \apjl, 813, L8 

\bibitem[Masters et al.(2012)]{Masters12} Masters, D., Capak, P., Salvato, M., et al.\ 2012, \apj, 755, 169

\bibitem[Matsuoka et al.(2018)]{Matsuoka18} Matsuoka, Y., Strauss, M.~A., Kashikawa, N., et al.\ 2018, \apj, 869, 150

\bibitem[Mazzucchelli et al.(2017)]{Mazzucchelli17} Mazzucchelli, C., Ba{\~n}ados, E., Venemans, B.~P., et al.\ 2017, \apj, 849, 91 

\bibitem[McGreer et al.(2018)]{McGreer18} McGreer, I.~D., Fan, X., Jiang, L., \& Cai, Z.\ 2018, \aj, 155, 131 

\bibitem[Mendel et al.(2014)]{Mendel14} Mendel, J.~T., Simard, L., Palmer, M., et al.\ 2014, \apjs, 210, 3

\bibitem[Ni et al.(2019)]{Ni19} Ni, Y., Di Matteo, T., Gilli, R., et al.\ 2020, \mnras, 495, 2135

\bibitem[Niida et al.(2020)]{Niida20} Niida, M., Nagao, T., Ikeda, H., et al.\ 2020, \apj, 904, 89

\bibitem[Onoue et al.(2019)]{Onoue19} Onoue, M., Kashikawa, N., Matsuoka, Y., et al.\ 2019, \apj, 880, 77

\bibitem[Pacucci \& Loeb(2020)]{Pacucci20} Pacucci, F. \& Loeb, A.\ 2020, \apj, 889, 52

\bibitem[Palanque-Delabrouille et al.(2016)]{Palanque16} Palanque-Delabrouille, N., Magneville, C., Y{\`e}che, C., et al.\ 2016, \aap, 587, A41

\bibitem[Ren et al.(2020)]{Ren20} Ren, K., Trenti, M., \& Di Matteo, T.\ 2020, \apj, 894, 124

\bibitem[Ross et al.(2013)]{Ross13} Ross, N.~P., McGreer, I.~D., White, M., et al.\ 2013, \apj, 773, 14

\bibitem[Schindler et al.(2019)]{Schindler19} Schindler, J.-T., Fan, X., McGreer, I.~D., et al.\ 2019, \apj, 871, 258

\bibitem[Schindler et al.(2020)]{Schindler20} Schindler, J.-T., Farina, E.~P., Ba{\~n}ados, E., et al.\ 2020, \apj, 905, 51

\bibitem[Shen et al.(2019)]{Shen19} Shen, Y., Wu, J., Jiang, L., et al.\ 2019, \apj, 873, 35

\bibitem[Shen et al.(2020)]{Shen20} Shen, X., Hopkins, P.~F., Faucher-Gigu{\`e}re, C.-A., et al.\ 2020, \mnras, 495, 3252

\bibitem[Shin et al.(2019)]{Shin19} Shin, J., Nagao, T., Woo, J.-H., et al.\ 2019, \apj, 874, 22

\bibitem[Shin et al.(2020)]{Shin20} Shin, S., Im, M., Kim, Y., et al.\ 2020, \apj, 893, 45

\bibitem[Tucci \& Volonteri(2017)]{Tucci17} Tucci, M. \& Volonteri, M.\ 2017, \aap, 600, A64

\bibitem[Veale et al.(2014)]{Veale14} Veale, M., White, M., \& Conroy, C.\ 2014, \mnras, 445, 1144

\bibitem[Vito et al.(2018)]{Vito18} Vito, F., Brandt, W.~N., Yang, G., et al.\ 2018, \mnras, 473, 2378

\bibitem[Willott et al.(2010)]{Willott10} Willott, C.~J., 
Delorme, P., Reyl{\'e}, C., et al.\ 2010, \aj, 139, 906 

\bibitem[Wyithe \& Loeb(2002)]{Wyithe02} Wyithe, J.~S.~B. \& Loeb, A.\ 2002, \apj, 577, 57

\bibitem[Yang et al.(2016)]{Yang16} Yang, J., Wang, F., Wu, X.-B., et al.\ 2016, \apj, 829, 33 

\bibitem[Yoon et al.(2019)]{Yoon19} Yoon, Y., Im, M., Hyun, M., et al.\ 2019, \apj, 871, 57

\end{thebibliography}
\end{document}